\documentclass[Physsubmission, Phys]{SciPost}
\usepackage{xspace}
\usepackage{placeins}

\newcommand{\Pj}{\ensuremath{\text{j}}\xspace}
\newcommand{\Pp}{\ensuremath{\text{p}}\xspace}

\newcommand{\Pd}{\ensuremath{\text{d}}\xspace}
\newcommand{\Ps}{\ensuremath{\text{s}}\xspace}
\newcommand{\Pc}{\ensuremath{\text{c}}\xspace}

\newcommand{\PW}{\ensuremath{\text{W}}\xspace}

\newcommand{\GeV}{\ensuremath{\,\text{GeV}}\xspace}
\newcommand{\TeV}{\ensuremath{\,\text{TeV}}\xspace}

\newcommand{\alphas}{\ensuremath{\alpha_\text{s}}\xspace}

\hypersetup{
    colorlinks,
    linkcolor={red!50!black},
    citecolor={blue!50!black},
    urlcolor={blue!80!black}
}

\usepackage[bitstream-charter]{mathdesign}
\DeclareSymbolFont{usualmathcal}{OMS}{cmsy}{m}{n}
\DeclareSymbolFontAlphabet{\mathcal}{usualmathcal}

\begin{document}

\begin{center}{\Large \textbf{
W+c-jet production at the LHC with NNLO QCD accuracy
}}\end{center}

\begin{center}
Micha\l{} Czakon\textsuperscript{1},
Alexander Mitov\textsuperscript{2},
Mathieu Pellen\textsuperscript{3$\star$}, and
Rene Poncelet\textsuperscript{2}
\end{center}

\begin{center}
{\bf 1} Institut f{\"u}r Theoretische Teilchenphysik und Kosmologie, RWTH Aachen University,\\
RWTH Aachen University, D-52056 Aachen, Germany
\\
{\bf 2} Cavendish Laboratory, University of Cambridge,\\
J.J.\ Thomson Avenue, Cambridge CB3 0HE, United Kingdom
\\
{\bf 3} Albert-Ludwigs-Universit\"at Freiburg, Physikalisches Institut,\\
Hermann-Herder-Stra\ss e 3, D-79104 Freiburg, Germany
\\
* mathieu.pellen@physik.uni-freiburg.de
\end{center}

\begin{center}
\today
\end{center}


\definecolor{palegray}{gray}{0.95}
\begin{center}
\colorbox{palegray}{
  \begin{tabular}{rr}
  \begin{minipage}{0.1\textwidth}
    \includegraphics[width=35mm]{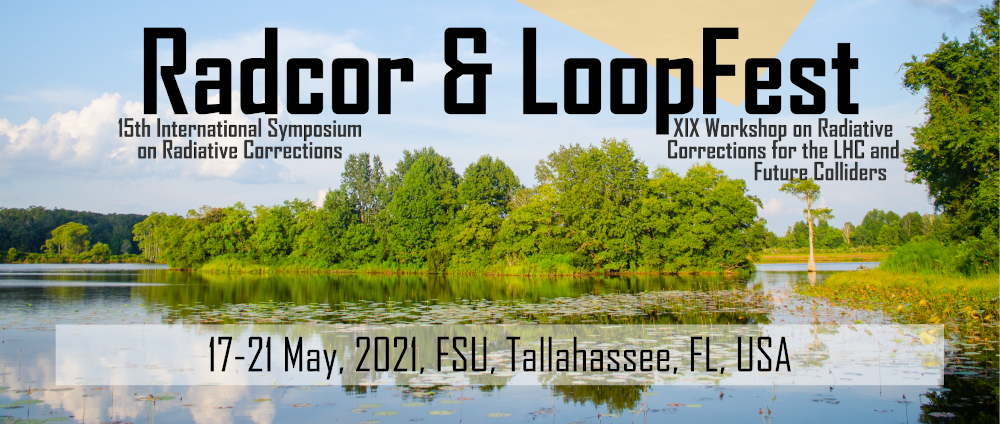}
  \end{minipage}
  &
  \begin{minipage}{0.85\textwidth}
    \begin{center}
    {\it 15th International Symposium on Radiative Corrections: \\Applications of Quantum Field Theory to Phenomenology,}\\
    {\it FSU, Tallahasse, FL, USA, 17-21 May 2021} \\
    \doi{10.21468/SciPostPhysProc.?}\\
    \end{center}
  \end{minipage}
\end{tabular}
}
\end{center}

\section*{Abstract}
{\bf
In these proceedings, we highlight some aspects of the recent computation of NNLO QCD corrections for W production in association with a charm jet at the LHC. 
The results are presented in the form of cross sections and differential distributions and are compared to ATLAS data.
}

\vspace{10pt}
\noindent\rule{\textwidth}{1pt}

\section{Introduction}
\label{sec:intro}

\begin{figure}
\center
                 \includegraphics[width=0.4\textwidth,page=1]{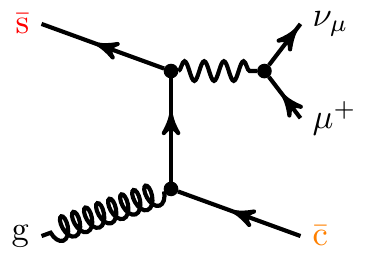}
                 \hspace{0.3cm}
                 \includegraphics[width=0.4\textwidth,page=1]{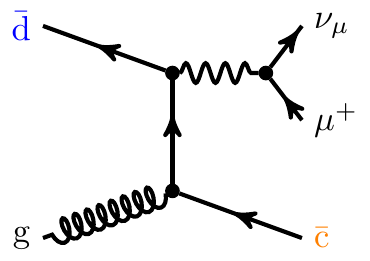}
        \caption{\label{fig:feynman}%
                Feynman diagrams of $\Pp\Pp \to \PW^+\Pj_{\rm c}$ at LO with diagonal CKM matrix (left) and off-diagonal elements (right).
                }
\end{figure}

The production of W boson in association with a charm jet at the LHC has raised significant interest in the recent years.
Beyond being interesting on its own as another test of perturbative QCD, the process is very sensitive to the parton distribution function (PDF) of the strange/anti-strange quark in the proton.
This is simply due to the direct link between W+c-jet production and strange/anti-strange quarks in the initial state as illustrated in the left-hand side of fig.~\ref{fig:feynman}.
It means that this process constitutes a key ingredient for the determination of the strange/anti-strange quark PDF \cite{Lai:2007dq,Faura:2020oom}.
To that end, several experimental analyses by both the ATLAS \cite{Aad:2014xca} and CMS collaborations \cite{CMS:2018muk,Sirunyan:2018hde,CMS:2019rlx} have been carried out.

On the theory side, next-to-leading order (NLO) QCD corrections have been known for quite some time for both the Tevatron \cite{Giele:1995kr} and the LHC \cite{Stirling:2012vh}.
Very recently, these corrections have been matched to parton shower \cite{Bevilacqua:2021ovq}.
In these proceedings, we briefly review some of the results of ref.~\cite{Czakon:2020coa}, where the first next-to-next-to-leading order (NNLO) QCD corrections to W production in association with a charm jet at the LHC have been computed.
In the original article, more details are provided such as predictions with different PDF sets, differential ratios between the ${\rm W^+}$ and ${\rm W^-}$ signatures etc.

The structure of the proceedings reads as follow: in section~\ref{sec:resuts}, our best prediction is compared to ATLAS data \cite{Aad:2014xca} at $7\TeV$ and the results are briefly discussed.
In section~\ref{sec:conclusion}, a short conclusion is provided.
For more information, the interested reader is referred to the original article \cite{Czakon:2020coa} as all the inputs utilised for the numerical simulations can be found there.

\section{Results}
\label{sec:resuts}

In the introduction, the direct link between strange PDF and W+charm production at the LHC has been explained.
Nonetheless, it is worth emphasising that this relation holds only at leading order (LO) when assuming a diagonal CKM matrix.
Considering off-diagonal CKM elements (see the right-hand side of fig.~\ref{fig:feynman}) or including higher-order QCD corrections significantly complicates the situation by adding new partonic channels.
This warrants therefore the precise computation of QCD corrections for this process.
For the predictions presented here, the effect of $V_{\Pc\Pd}\neq0$ is included at Born level only.
Also, all theoretical predictions presented here have been obtained withing the {\sc Stripper} framework which is a c++ implementation of the four-dimensional formulation of the sector-improved residue subtraction scheme \cite{Czakon:2010td,Czakon:2011ve,Czakon:2014oma,Czakon:2019tmo}.

Following the ATLAS analysis \cite{Aad:2014xca}, the phase-space definition reads
\begin{align}
 p_{\mathrm{T},\ell} >  20\GeV, \qquad |\eta_\ell| < 2.5, \qquad
 p_{\mathrm{T},\text{miss}} >  25\GeV, \qquad m_{\rm T}^{\rm W} > 40\GeV,
\end{align}
for the leptonic final states.
In addition, one and only one charm jet should fulfil
\begin{align}
 p_{\mathrm{T},\Pj_c} >  25\GeV, \qquad |\eta_{\Pj_c}| < 2.5.
\end{align}

\begin{figure}[h!]
        \setlength{\parskip}{-10pt}
                \hspace{-10pt}
                 \includegraphics[width=\textwidth,page=1]{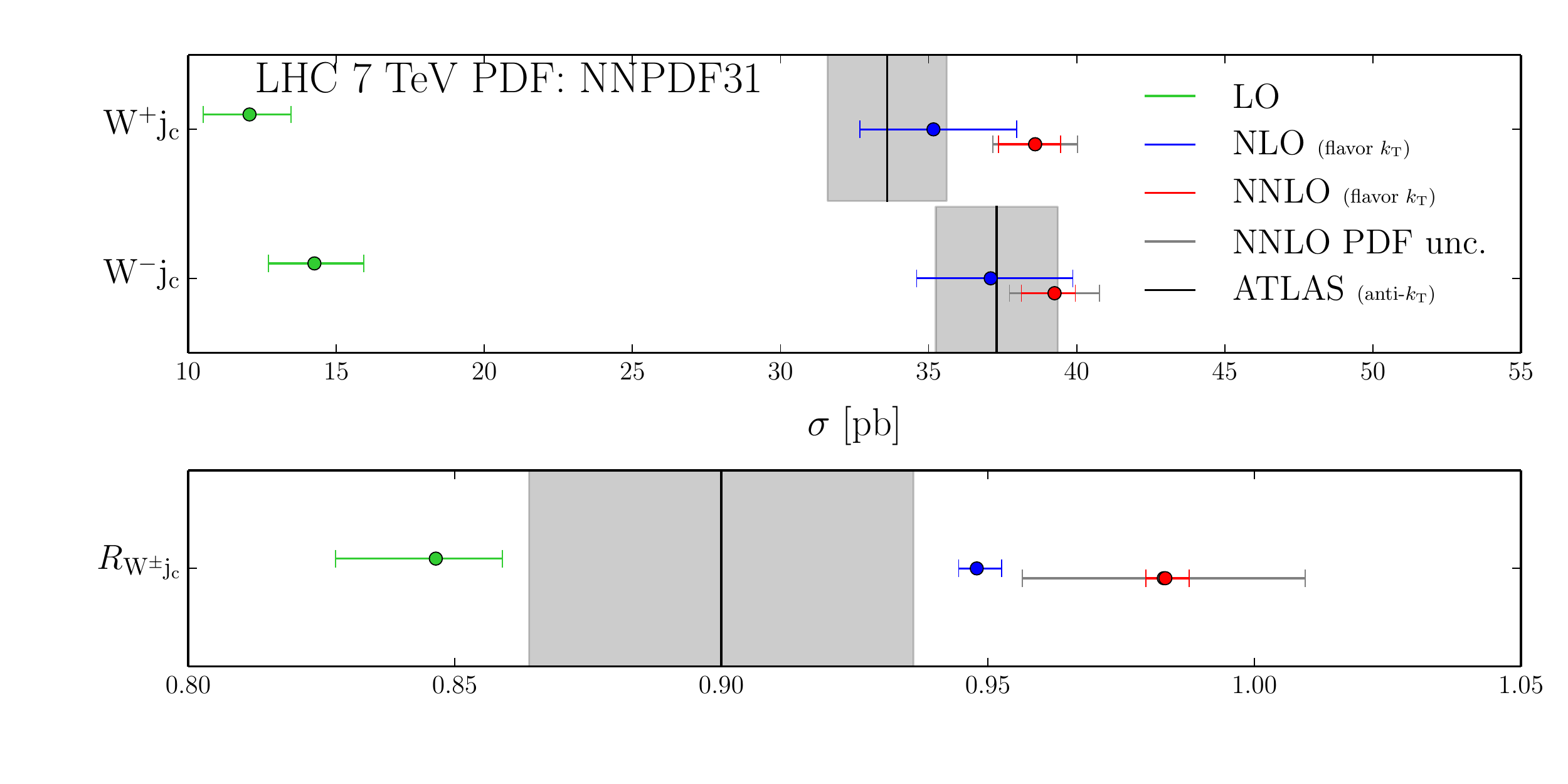}
        \vspace*{-3ex}
        \caption{\label{fig:xsection}%
                Cross sections for $\Pp\Pp \to \PW^+\Pj_{\rm c}$, $\Pp\Pp \to \PW^-\Pj_{\rm c}$, and the ratio $R_{\PW^\pm\Pj_{\rm c}}$ at the LHC with $\sqrt{s}=7\TeV$.
                The theoretical predictions up to NNLO QCD accuracy are compared to the ATLAS data \cite{Aad:2014xca}.}
\end{figure}

In fig.~\ref{fig:xsection}, the cross sections of $\Pp\Pp \to \PW^+\Pj_{\rm c}$ and $\Pp\Pp \to \PW^-\Pj_{\rm c}$ are compared to the measurements of the ATLAS collaboration \cite{Aad:2014xca}.
In addition, the ratio of the two cross sections $R_{\PW^\pm\Pj_{\rm c}} = \sigma_{\PW^+\Pj_{\rm c}}/\sigma_{\PW^-\Pj_{\rm c}}$ is also provided.
Based on the previous discussion, this ratio behaves approximately like $R_{\PW^\pm\Pj_{\rm c}} \sim$ 
$\left(|V_{\Pc\Ps}|^2\bar\Ps +|V_{\Pc\Pd}|^2\bar \Pd \right)/\left(|V_{\Pc\Ps}|^2\Ps +|V_{\Pc\Pd}|^2\Pd \right)$,
 meaning that it also provides a sensitive probe of the strange-quark PDF.

 The first interesting aspect to observe is that NLO QCD corrections are large while the NNLO ones are much more modest. This is by now a very well understood phenomenon for V+j processes \cite{Rubin:2010xp} and is also a sign of good perturbative convergence.
 As expected, the scale uncertainty is significantly decreasing when including higher orders.
 At NNLO, it becomes of the order of 2-3 per cent, 
 meaning that it is smaller than the uncertainty due to PDF variation which is around $4\%$.
 Finally, the NNLO computations and the experimental data agree within their respective uncertainties. This holds true for both signatures as well as the ratio.

\begin{figure}[h!]
                 \includegraphics[width=0.5\textwidth,page=1]{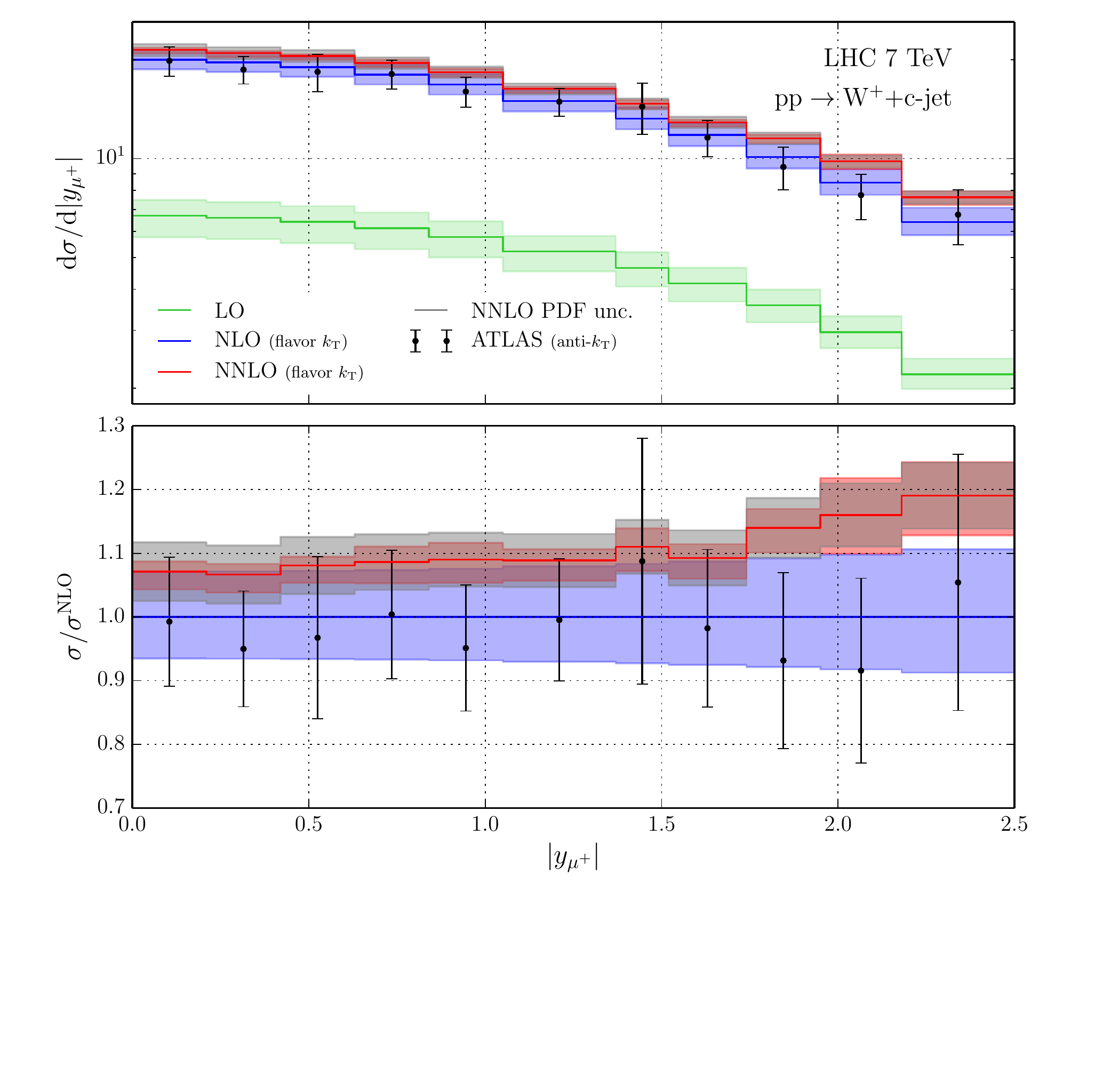}
                 \includegraphics[width=0.5\textwidth,page=1]{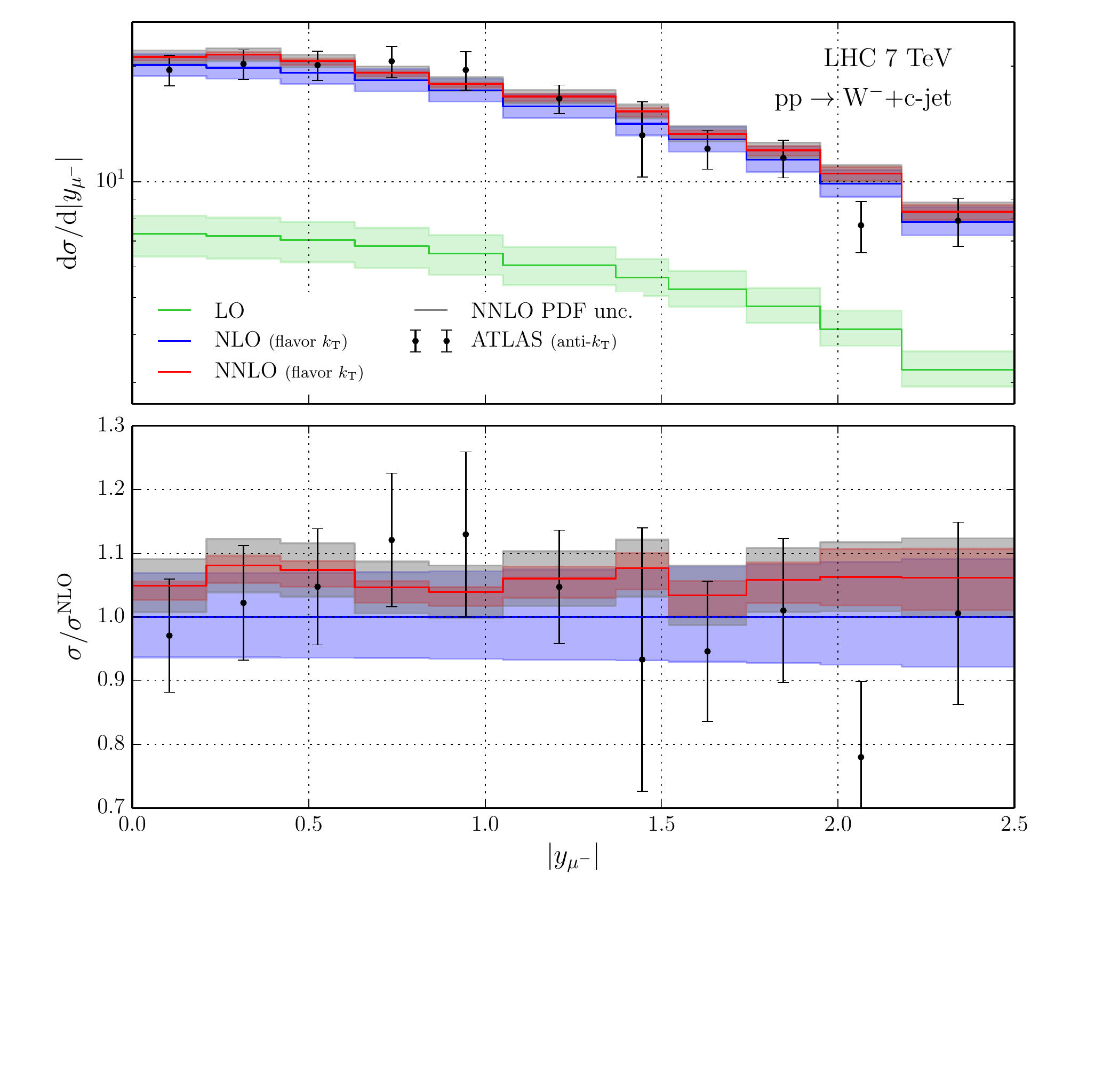}
        \vspace*{-9ex}
        \caption{\label{fig:data}%
                Differential distributions in the absolute rapidity of the muon/anti-muon in the process 
                $\Pp\Pp \to \PW^+\Pj_{\rm c}$ (left) and of the muon in $\Pp\Pp \to \PW^-\Pj_{\rm c}$ (right) at the LHC with $\sqrt{s}=7\TeV$.
                The upper panel shows the absolute predictions as well as the ATLAS data \cite{Aad:2014xca}.}
\end{figure}

At the level of the differential distributions, the general picture is the same, with large NLO corrections and relatively moderate NNLO effects across the phase space.
As an example, the differential distribution in the absolute rapidity of the muon/anti-muon is given in fig.~\ref{fig:data}. It is shown for both processes $\Pp\Pp \to \PW^+\Pj_{\rm c}$ (left) and $\Pp\Pp \to \PW^-\Pj_{\rm c}$ (right).
The uncertainty due to PDF variation is again larger than the scale uncertainty obtained at NNLO accuracy.
In the same way as for the fiducial cross section, the data-theory agreement is rather good over the whole kinematic range.

Nonetheless, it is worth mentioning a few limitations of this comparison.
First, in our computation we have made use of the flavor-$k_{\rm T}$ algorithm \cite{Banfi:2006hf} to ensure an infrared safe definition of the charm jets.
On the other hand, the ATLAS measurement has been performed using the anti-$k_{\rm T}$ algorithm \cite{Cacciari:2008gp}.
For the case of ${\rm Z+b}$ production, this mismatch has been estimated to be around $10\%$ \cite{Gauld:2020deh}.
In the future, it is thus worth investigating such effect for W+c measurements.
Second, the effects of $V_{\Pc\Pd} \neq 0$ has only been included at LO in our computation.
At this order, it amounts to $5\%$ to $10\%$ depending on the signature.
It is thus expected that such effects should amount to few per cent at higher orders.
Finally, electroweak (EW) corrections have been here neglected.
Due to Sudakov logarithms, these are usually around few per cent and grow negatively in high-energy limits.
The EW corrections of order $\mathcal{O}\left(\alphas\alpha^3\right)$ have been found to be around $-3\%$ \cite{Denner:2009gj} while the subleading ones of order $\mathcal{O}\left(\alpha^4\right)$ are expected to be below a per cent \cite{Denner:2019zfp}.

\section{Conclusion}
\label{sec:conclusion}

In these proceedings we have reported on a recent computations of NNLO QCD effects for the W+c-jet production at the LHC.
In particular, we have focused on the main highlight of ref.~\cite{Czakon:2020coa} which is the comparison of our best predictions with the ATLAS measurement of W+c-jet at $7\TeV$ \cite{Aad:2014xca}.
Overall the agreement between theory and data is good.
Nonetheless in order to perform precision comparisons, several aspects should be addressed in the future: the inclusion of off-diagonal CKM elements at higher order, the inclusion of EW corrections, and finally quantifying the effect of different jet algorithms for the charm jets identification.
This work constitutes therefore only a first step toward determining with high precision the strange-quark content of the proton.

\section*{Acknowledgements}

The work of M.C. was supported by the Deutsche Forschungsgemeinschaft under grant 396021762 -- TRR 257.
The research of A.M., M.P., and R.P. has received funding from the European Research Council (ERC) under the European Union's Horizon 2020 Research and Innovation Programme (grant agreement no.~683211).
A.M. was also supported by the UK STFC grants ST/L002760/1 and ST/K004883/1.
M.P. acknowledges support by the German Research Foundation (DFG) through the Research Training Group RTG2044.
R.P. acknowledges support from the Leverhulme Trust and the Isaac Newton Trust.

\bibliographystyle{SciPost_bibstyle} 
\bibliography{Wc.bib}

\begin{thebibliography}{10}
\providecommand{\url}[1]{\texttt{#1}}
\providecommand{\urlprefix}{URL }
\expandafter\ifx\csname urlstyle\endcsname\relax
  \providecommand{\doi}[1]{doi:\discretionary{}{}{}#1}\else
  \providecommand{\doi}{doi:\discretionary{}{}{}\begingroup
  \urlstyle{rm}\Url}\fi
\providecommand{\eprint}[2][]{\url{#2}}

\bibitem{Lai:2007dq}
H.~L. Lai, P.~M. Nadolsky, J.~Pumplin, D.~Stump, W.~K. Tung and C.~P. Yuan,
\newblock \emph{{The Strange parton distribution of the nucleon: Global
  analysis and applications}},
\newblock JHEP \textbf{04}, 089 (2007),
\newblock \doi{10.1088/1126-6708/2007/04/089},
\newblock \eprint{hep-ph/0702268}.

\bibitem{Faura:2020oom}
F.~Faura, S.~Iranipour, E.~R. Nocera, J.~Rojo and M.~Ubiali,
\newblock \emph{{The Strangest Proton?}},
\newblock Eur. Phys. J. C \textbf{80}(12), 1168 (2020),
\newblock \doi{10.1140/epjc/s10052-020-08749-3},
\newblock \eprint{2009.00014}.

\bibitem{Aad:2014xca}
G.~Aad \emph{et~al.},
\newblock \emph{{Measurement of the production of a $W$ boson in association
  with a charm quark in $pp$ collisions at $\sqrt{s} =$ 7 TeV with the ATLAS
  detector}},
\newblock JHEP \textbf{05}, 068 (2014),
\newblock \doi{10.1007/JHEP05(2014)068},
\newblock \eprint{1402.6263}.

\bibitem{CMS:2018muk}
C.~Collaboration,
\newblock \emph{{Measurement of associated production of W bosons with charm
  quarks in proton-proton collisions at $\sqrt{s}=13~\mathrm{TeV}$ with the CMS
  experiment at the LHC}}  (2018).

\bibitem{Sirunyan:2018hde}
A.~M. Sirunyan \emph{et~al.},
\newblock \emph{{Measurement of associated production of a W boson and a charm
  quark in proton-proton collisions at $\sqrt{s} =$ 13 TeV}},
\newblock Eur. Phys. J. \textbf{C79}(3), 269 (2019),
\newblock \doi{10.1140/epjc/s10052-019-6752-1},
\newblock \eprint{1811.10021}.

\bibitem{CMS:2019rlx}
C.~Collaboration,
\newblock \emph{{Measurement of the associated production of a W boson and a
  charm quark at $\sqrt{s}=8~\mathrm{TeV}$}}  (2019).

\bibitem{Giele:1995kr}
W.~T. Giele, S.~Keller and E.~Laenen,
\newblock \emph{{QCD corrections to $W$ boson plus heavy quark production at
  the Tevatron}},
\newblock Phys. Lett. B \textbf{372}, 141 (1996),
\newblock \doi{10.1016/0370-2693(96)00078-0},
\newblock \eprint{hep-ph/9511449}.

\bibitem{Stirling:2012vh}
W.~Stirling and E.~Vryonidou,
\newblock \emph{{Charm production in association with an electroweak gauge
  boson at the LHC}},
\newblock Phys. Rev. Lett. \textbf{109}, 082002 (2012),
\newblock \doi{10.1103/PhysRevLett.109.082002},
\newblock \eprint{1203.6781}.

\bibitem{Bevilacqua:2021ovq}
G.~Bevilacqua, M.~V. Garzelli, A.~Kardos and L.~Toth,
\newblock \emph{{W+charm production with massive c quarks in PowHel}}  (2021),
\newblock \eprint{2106.11261}.

\bibitem{Czakon:2020coa}
M.~Czakon, A.~Mitov, M.~Pellen and R.~Poncelet,
\newblock \emph{{NNLO QCD predictions for W+c-jet production at the LHC}},
\newblock JHEP \textbf{06}, 100 (2021),
\newblock \doi{10.1007/JHEP06(2021)100},
\newblock \eprint{2011.01011}.

\bibitem{Czakon:2010td}
M.~Czakon,
\newblock \emph{{A novel subtraction scheme for double-real radiation at
  NNLO}},
\newblock Phys. Lett. B \textbf{693}, 259 (2010),
\newblock \doi{10.1016/j.physletb.2010.08.036},
\newblock \eprint{1005.0274}.

\bibitem{Czakon:2011ve}
M.~Czakon,
\newblock \emph{{Double-real radiation in hadronic top quark pair production as
  a proof of a certain concept}},
\newblock Nucl. Phys. B \textbf{849}, 250 (2011),
\newblock \doi{10.1016/j.nuclphysb.2011.03.020},
\newblock \eprint{1101.0642}.

\bibitem{Czakon:2014oma}
M.~Czakon and D.~Heymes,
\newblock \emph{{Four-dimensional formulation of the sector-improved residue
  subtraction scheme}},
\newblock Nucl. Phys. B \textbf{890}, 152 (2014),
\newblock \doi{10.1016/j.nuclphysb.2014.11.006},
\newblock \eprint{1408.2500}.

\bibitem{Czakon:2019tmo}
M.~Czakon, A.~van Hameren, A.~Mitov and R.~Poncelet,
\newblock \emph{{Single-jet inclusive rates with exact color at $ \mathcal{O} $
  ($ {\alpha}_s^4 $)}},
\newblock JHEP \textbf{10}, 262 (2019),
\newblock \doi{10.1007/JHEP10(2019)262},
\newblock \eprint{1907.12911}.

\bibitem{Rubin:2010xp}
M.~Rubin, G.~P. Salam and S.~Sapeta,
\newblock \emph{{Giant QCD K-factors beyond NLO}},
\newblock JHEP \textbf{09}, 084 (2010),
\newblock \doi{10.1007/JHEP09(2010)084},
\newblock \eprint{1006.2144}.

\bibitem{Banfi:2006hf}
A.~Banfi, G.~P. Salam and G.~Zanderighi,
\newblock \emph{{Infrared safe definition of jet flavor}},
\newblock Eur. Phys. J. \textbf{C47}, 113 (2006),
\newblock \doi{10.1140/epjc/s2006-02552-4},
\newblock \eprint{hep-ph/0601139}.

\bibitem{Cacciari:2008gp}
M.~Cacciari, G.~P. Salam and G.~Soyez,
\newblock \emph{{The anti-$k_t$ jet clustering algorithm}},
\newblock JHEP \textbf{04}, 063 (2008),
\newblock \doi{10.1088/1126-6708/2008/04/063},
\newblock \eprint{0802.1189}.

\bibitem{Gauld:2020deh}
R.~Gauld, A.~Gehrmann-De~Ridder, E.~W.~N. Glover, A.~Huss and I.~Majer,
\newblock \emph{{Predictions for $\mathrm{Z}$-boson production in association
  with a $\mathrm{b}$-jet at $\mathcal{O}(\alpha_s^3)$}}  (2020),
\newblock \eprint{2005.03016}.

\bibitem{Denner:2009gj}
A.~Denner, S.~Dittmaier, T.~Kasprzik and A.~M{\"u}ck,
\newblock \emph{{Electroweak corrections to W+jet hadroproduction including
  leptonic W-boson decays}},
\newblock JHEP \textbf{08}, 075 (2009),
\newblock \doi{10.1088/1126-6708/2009/08/075},
\newblock \eprint{0906.1656}.

\bibitem{Denner:2019zfp}
A.~Denner, S.~Dittmaier, M.~Pellen and C.~Schwan,
\newblock \emph{{Low-virtuality photon transitions $\gamma^*\to f\bar f$ and
  the photon-to-jet conversion function}},
\newblock Phys. Lett. \textbf{B798}, 134951 (2019),
\newblock \doi{10.1016/j.physletb.2019.134951},
\newblock \eprint{1907.02366}.

\end{thebibliography}

\nolinenumbers

\end{document}